\documentclass{PoS}
\usepackage{cite}

\newcommand{ \be }{\begin{equation}}
\newcommand{ \ee }{\end{equation}}

\newcommand{\sNN}{{{$\sqrt{s_{_{{\mathrm{NN}}}}}$}}}

\newcommand{\muB}{\mbox{$\mu_{B}$}}

\newcommand{\KV}{{\mbox{$\kappa\sigma^{2}$}}}
\newcommand{\SD}{{\mbox{$S\sigma$}}}

\title{Energy Dependence of Moments of Net-Proton and Net-Charge Multiplicity Distributions at STAR}

\ShortTitle{Moments of Net-Proton and Net-Charge Multiplicity Distributions at STAR}

\author{\speaker{Xiaofeng Luo}  (for the STAR Collaboration)\\
        Institute of Particle Physics and Key Laboratory of Quark \& Lepton Physics (MOE)
        , Central China Normal University, Wuhan, 430079, China.\\
        E-mail: \email{xfluo@mail.ccnu.edu.cn}}

\abstract{ We present the energy dependence of moments of net-proton and net-charge multiplicity distributions in Au+Au collisions measured by the STAR experiment in the first phase of the Beam Energy Scan (BES) at the Relativistic Heavy Ion Collider (RHIC). By using the time of flight detector for particle identification, the upper transverse momentum ($p_{T}$) limit for proton and anti-proton is extended from 0.8 GeV/c up to 2 GeV/c. The $p_{T}$ and rapidity acceptance dependence study for the moments of net-proton distribution show that the larger the acceptance is, the greater the deviation from unity. The most pronounced structure is found in the energy dependence of {\KV} of net-proton distributions from the $0\sim5\%$ most central collisions within $0.4<p_{T}<2$ GeV/c and at mid-rapidity $|y|<0.5$. At energies above 39 GeV,  the values of {\KV} are close to unity and for energies below 39 GeV, it shows significant deviation below unity around 19.6 and 27 GeV, then a large increase above unity is observed at 7.7 GeV.

 }

\FullConference{9th International Workshop on Critical Point and Onset of Deconfinement - CPOD2014,\\
		17-21 November 2014\\
		ZiF (Center of Interdisciplinary Research), University of Bielefeld, Germany}

\begin{document}

\section{Introduction}
The phase structure of strongly interacting, hot and dense nuclear matter can be illustrated in the two dimensional Quantum Chromodynamics (QCD)
phase diagram (temperature, $\mathrm{T}$ vs. baryon chemical potential, $\mu_{B}$). First principle Lattice QCD calculations show a smooth crossover~\cite{crossover}
from the hadronic to the Quark Gluon Plasma (QGP) phase at vanishing baryon chemical potential ($\mu_{B}=0$) and finite temperature. At large $\mu_{B}$, theoretical calculations demonstrate that the phase transition is of the first-order~\cite{firstorder}. Thus, there should be a so-called QCD Critical Point (CP) at the end point of the first-order phase transition boundary towards the crossover region~\cite{QCP_Prediction}. Due to the sign problem of Lattice QCD at finite {\muB}, there is large uncertainty in determining the location of the CP 
or even its existence~\cite{qcp,qcp_Rajiv}. Experimental confirmation of the existence of the CP will be an excellent verification of QCD theory in the non-perturbative region and a milestone of the exploration of the QCD phase structure.
It is one of the main goals of the Beam Energy Scan (BES) program at the Relativistic Heavy Ion Collider (RHIC). By tuning the colliding energy of gold nuclei from high to low values, one can vary the temperature and baryon chemical 
potential of the hot, dense nuclear matter created in high energy nuclear collisions~\cite{bes}. This can provide us a unique experimental tool to probe the QCD phase structure in a broad $T$ and $\muB$ region. Fluctuations of conserved quantities, such as 
net-baryon (B), net-charge (Q) and net-strangeness (S),  can serve as powerful observables to probe the QCD phase transition and CP signal in heavy-ion collisions.  These observables are sensitive to the correlation length of the system~\cite{qcp_signal,ratioCumulant,Neg_Kurtosis}  and directly connected to the susceptibilities in the theoretical calculations~\cite{science,Lattice}. Theoretically, net-proton number 
fluctuation is a reasonable  proxy of the fluctuation of net-baryon number in measuring critical fluctuations near the QCD critical point~\cite{Hatta}. 
However, if one considers the effects of the non-critical contributions associated with heavy-ion collision dynamics, such as hadronic scattering, resonance decay and baryon number conservation, 
there would be differences between net-proton and net-baryon fluctuations~\cite{Asakawa_formula,QM2014_baseline}. To understand those non-critical contributions to the observables in terms of searching for the CP, careful studies are necessary.

In this paper, we present new analysis of the moments of net-proton distributions within the wider transverse momentum range 
$0.4<p_{T}<2$ GeV/c in Au+Au collisions at {\sNN} = 7.7 , 11.5, 19.6, 27, 39, 62.4 and 200 GeV. The paper is organized as follows: In the second section, we will give a brief discussion of
the techniques used in the moment analysis; the experimental results for moments of net-proton and net-charge distributions will be
presented in the third section; last, we will give a summary and outlook.

\section{Analysis Techniques}
 In heavy-ion collisions, one cannot directly measure the collision centrality and/or initial collision geometry of two nuclei. The centrality in heavy-ion collision experiments is generally determined by comparing the measured particle multiplicity with the Monte Carlo Glauber simulations. The centrality is usually denoted as a percentage (e.g. 0-5\%, 5-10\%,...) over a collection of events to represent the fraction of the total cross section. Generally, this can result in two undesirable effects in the moment analysis of particle multiplicity distributions within finite centrality bins. One is the so-called centrality bin width effect, which is caused by volume variation within a finite centrality bin size;  the other one is the centrality resolution effect, which is due to the initial volume fluctuation. During last five years,  several analysis techniques have been used in the moments analysis to address those background effects and extract the dynamical fluctuation signals from the observables in heavy-ion collisions.
Those include
: (1) Centrality bin width correction~\cite{WWND2011,technique}, (2) Novel centrality determination to account for the effects of centrality resolution and auto-correlation~\cite{technique}, and  (3) Efficiency correction and error estimation~\cite{Delta_theory,voker_eff1,voker_eff2,Unified_Errors}. Let us briefly discuss those techniques in the following sub-sections.
\subsection{Centrality Bin Width Correction}
The centrality bin width correction addresses the volume fluctuation effects on the moments of multiplicity distributions within a finite width centrality bin. It is also called the centrality bin width effect. This effect needs to be eliminated, as an artificial centrality dependence could be introduced due to this effect. To do this, the
centrality bin width correction is applied to calculate the various moments of particle multiplicity
distributions in one wide centrality bin. Experimentally, the smallest centrality bin is determined by a
single value of particle multiplicity.
Experimental results are usually reported for a wider centrality bin (a range of particle multiplicity),
such as $0-5\%$,$5-10\%$,...etc., to reduce statistical errors. To eliminate the centrality bin width effect, we
calculate the various order cumulants ($C_{n}$) for each single particle multiplicity within one
wider centrality bin and weigh the average by the corresponding number of events in that multiplicity:
\begin{equation} \label{eq:cbwc}
{C_n} = \frac{{\sum\limits_{r = {N_1}}^{{N_2}} {{n_r}C_n^r} }}{{\sum\limits_{r = {N_1}}^{{N_2}} {{n_r}} }} = \sum\limits_{r = {N_1}}^{{N_2}} {{\omega _r}C_n^r} 
\end{equation}
where the $n_r$ is the number of events for multiplicity value $r$ and
the corresponding weight for the multiplicity $r$, ${\omega _r} = {n_r}/\sum\limits_{r = {N_1}}^{{N_2}} {{n_r}}$. $N_1$ and $N_2$ are the lowest and highest multiplicity values for one centrality bin. Once the cumulants are corrected via Eq. (\ref{eq:cbwc}), we can calculate the various order moments products, for e.g. {\KV}$=C_{4}/C_{2}$ and {\SD}$=C_{3}/C_{2}$, where $\kappa$ and $S$ are kurtosis and skewness, respectively. The final statistical error of the moments for one centrality can be evaluated by standard error propagation based on Eq. (\ref{eq:cbwc}) from the errors of the finer bins.  For more details, see~\cite{technique}, and for details on the error calculation for the finer bins, see Sec. \ref{Eff_error} and ~\cite{Unified_Errors}.

\subsection{Centrality Resolution and Auto-correlation Effects}
Particle multiplicity are typically used to determine centrality since it can reflect the initial geometry of heavy-ion collision. However, the relation
between measured particle multiplicities and collision geometry is not one-to-one correspondence and there are 
fluctuations in the particle multiplicity even for a fixed collision geometry. Thus,
one could have a finite resolution of initial collision geometry (centrality resolution) by using particle multiplicity to determine the centrality.
The more particles that are used in the centrality determination, the better the centrality resolution and the smaller 
the fluctuation of the initial geometry (volume fluctuation). This may affect moments of the event-by-event multiplicity distributions.
In addition,  we need to avoid auto-correlation effects caused by the particles involved in the analysis also being used to calculate centrality.
The centrality resolution and auto-correlation effects can be well addressed by a novel centrality definition. For the net-proton moment analysis, the multiplicity of charged kaon and pion within pseudo-rapidity range $|\eta|<1$ is used to define the collision centrality
in Au+Au collisions. For net-charge, the collision centrality is defined by the multiplicity of charged particles within pseudo-rapidity range $0.5<|\eta|<1$.
For more details, one can see~\cite{technique,CPOD2013}.
\subsection{Efficiency Correction and Error Estimation} \label{Eff_error}
Experimentally, particle detectors have a finite particle detection efficiency due to the limited capability of the detector to register the incoming
particles. This efficiency effect will lead to loss of the measured multiplicity in every event, which will
change the shape of the original multiplicity distributions. Since higher
moments are very sensitive to the shape of the distributions, especially the tails, the moments
values could be significantly modified by the detector efficiency, which could distort and/or
suppress the original signal induced by the QCD critical point. To obtain precise moment
measurements for the QCD critical point search in heavy-ion collisions, it is very important
to recover the moments of the original multiplicity distributions with the measured ones by
applying an efficiency correction technique. The efficiency correction is not only important for the
values of the moments but also for the statistical errors. Since the fluctuation analysis is statistics hungry, it is crucial to get the correct statistical errors with limited statistics. 
In the paper~\cite{Unified_Errors}, we provide a unified description of efficiency correction and error estimation for various order moments of multiplicity distributions. The basic idea is to express the moments and cumulants in terms of the factorial moments, which can be easily corrected for efficiency effects. By knowing the covariance between multivariate factorial moments, we use the standard error propagation based on the Delta theorem to obtain the error formula for efficiency-corrected moments. This method can also be applied to the phase space efficiency case, where the efficiency of protons or anti-protons is not constant within the studied phase space. One needs to note that the efficiency correction and error estimation should be done just before the centrality bin width correction.

\section{Results and Discussion}
The STAR Collaboration has published the energy dependence of the moments of net-proton (proton number minus anti-proton number)~\cite{STAR_BES_PRL} and net-charge~\cite{netcharge_PRL}  multiplicity distributions in Au+Au collisions at ${\sqrt{s_{\mathrm{ NN}}}}$ = 7.7 , 11.5, 19.6, 27, 39, 62.4 and 200 GeV.
Those data are taken from the first phase of the RHIC BES program in the years 2010 and 2011.  In addition, the data of 14.5 GeV was successfully taken in the year 2014, which can fill in the large gap between 11.5 and 19.6 GeV.
For the net-charge moment analysis, the charged particles 
are measured within the transverse momentum range  $0.2<p_{T}<2$ GeV/c and pseudo-rapidity range $|\eta|<0.5$.
For the net-proton case, the protons and anti-protons are identified with ionization energy loss in the Time Projection
Chamber (TPC) of the STAR detector within transverse momentum range $0.4<p_{T}<0.8$ GeV/c and at mid-rapidity $|y|<0.5$. 
 
In this paper, we will present a new analysis of moments of the net-proton distributions  in Au+Au collisions at {\sNN} = 7.7 , 11.5, 19.6, 27, 39, 62.4 and 200 GeV. 
In this analysis, we extended the upper $p_{T}$ coverage for proton/anti-proton up to  2 GeV/c utilizing the time of flight (ToF) detector at STAR for particle identification. This would allow us to have more protons and anti-protons (about 2 times) in the net-proton moment analysis. The published net-charge results will be also discussed.

\begin{figure}[htbp]
%\begin{center}
\hspace{-0.8cm}
\vspace{-0.8cm}
\includegraphics[scale=0.8]{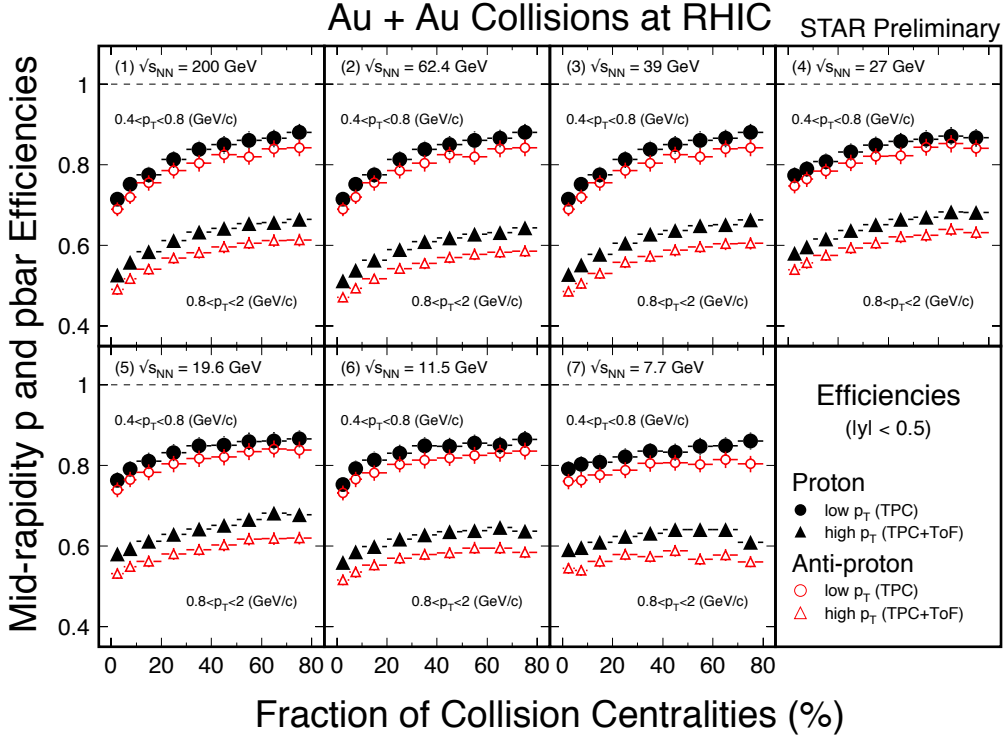}
\vspace{-1.2cm}
\caption[]{(Color online) Centrality dependence of mid-rapidity detecting efficiency for protons and anti-protons in two $p_{T}$ ranges, $0.4<p_{T}<0.8$ GeV/c (circles) and  $0.8<p_{T}<2$ GeV/c (triangles), in 
Au+Au collisions at {\sNN}=7.7 , 11.5, 19.6, 27, 39, 62.4 and 200 GeV. Black solid points represent efficiency of protons and red empty points are the efficiency of anti-protons.  } \label{fig:peff_BES}
%\end{center}
\end{figure}
Figure \ref{fig:peff_BES} shows the centrality dependence of detection efficiency for protons and anti-protons in two $p_{T}$ ranges, $0.4<p_{T}<0.8$ GeV/c (circles) and $0.8<p_{T}<2$ GeV/c (triangles),  in 
Au+Au collisions at {\sNN}=7.7 , 11.5, 19.6, 27, 39, 62.4 and 200 GeV. The efficiency of proton and anti-proton at high $p_{T}$, $0.8<p_{T}<2$ GeV/c is significantly lower than that of low $p_{T}$, $0.4<p_{T}<8$ GeV/c. This is a result of the ToF matching efficiency from the ToF detector at high $p_{T}$. At low $p_{T}$, only the TPC is used to identify protons and anti-protons. Thus, the total efficiency for protons and anti-protons at low $p_{T}$ can be obtained as Eff( $0.4<p_{T}<0.8$)=Eff(TPC, $0.4<p_{T}<0.8$ ) and  efficiency at high $p_{T}$ is calculated as Eff( $0.8<p_{T}<2$)=Eff(TPC,  $0.8<p_{T}<2$ )*Eff(ToF, $0.8<p_{T}<2$). The TPC efficiency of protons and anti-protons are obtained from the so-called embedding simulation technique and the ToF matching efficiency can be calculated from the data. The efficiency increases from central to peripheral collisions for all energies.  Due to material absorption of anti-protons, the efficiency of anti-protons is slightly lower than that of protons. Those efficiency numbers in Fig. \ref{fig:peff_BES} will be used in the efficiency correction formulae for the moments of the net-proton distributions
as discussed in ~\cite{Unified_Errors}.
\begin{figure}[htbp]
%\begin{center}
\hspace{-0.4in}
%\vspace{-0.8cm}
\includegraphics[scale=0.85]{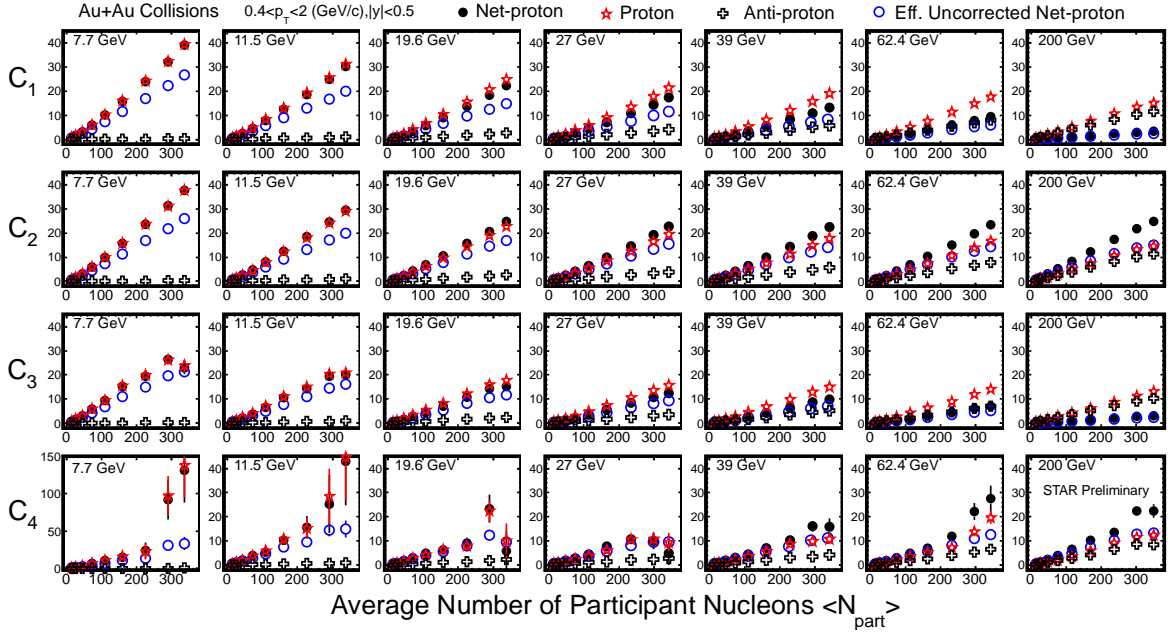}
%\vspace{-1.2cm}
\caption[]{(Color online) Centrality dependence of various order efficiency corrected cumulants  ($C_{1}\sim C_{4}$) for net-proton, proton and anti-proton distributions in 
Au+Au collisions at {\sNN}=7.7 , 11.5, 19.6, 27, 39, 62.4 and 200 GeV. Error bars in the figure are statistical errors only.   Blue empty circles represent the efficiency uncorrected cumulants of net-proton distributions. } \label{fig:Cumulants}
%\end{center}
\end{figure}

Figure \ref{fig:Cumulants} shows the centrality dependence of efficiency corrected cumulants ($C_{1}\sim C_{4}$) of net-proton, proton and anti-proton distributions in Au+Au collisions at {\sNN}=7.7 , 11.5, 19.6, 27, 39, 62.4 and 200 GeV. The protons and anti-protons are measured within transverse momentum $0.4<p_{T}<2$ GeV/c and at mid-rapidity ($|y|<0.5$).
Generally, various order cumulants of net-proton, proton and anti-proton distributions show a linear dependence on the average number of participant nucleons ($<N_{part}>$). The proton cumulants are always larger than the anti-proton cumulants and the difference between proton and anti-proton cumulants are larger in low energies than in high energies. 
The cumulants of net-proton distributions closely follow the proton cumulants when the colliding energy decreases. These observations can be explained as the interplay between the baryon stopping and 
pair production of protons and anti-protons. At high energies,  protons and anti-protons mainly come from the pair production and the number of protons and anti-protons are very similar. At low energies,  the production of protons is dominated by initial baryon stopping and the number of protons is far higher than the number of anti-protons. The values of the forth order cumulant ($C_{4}$) at 7.7 and 11.5 GeV significantly increase in the $0\sim5\%$ and $5\sim10\%$ centrality bins with respect to the efficiency uncorrected results. The efficiency correction not only affects the values but also lead to larger statistical errors, as $error(C_{n})\sim \sigma^{n}/\varepsilon^{\alpha}$, where the $\sigma$ in numerator is the standard deviation of the particle distributions and the denominator $\varepsilon$ is the efficiency number,  $\alpha$ is a positive real number~\cite{Unified_Errors}.
\begin{figure}[htbp]
%\begin{center}
\hspace{-0.7cm}
%\vspace{-0.8cm}
\includegraphics[scale=0.8]{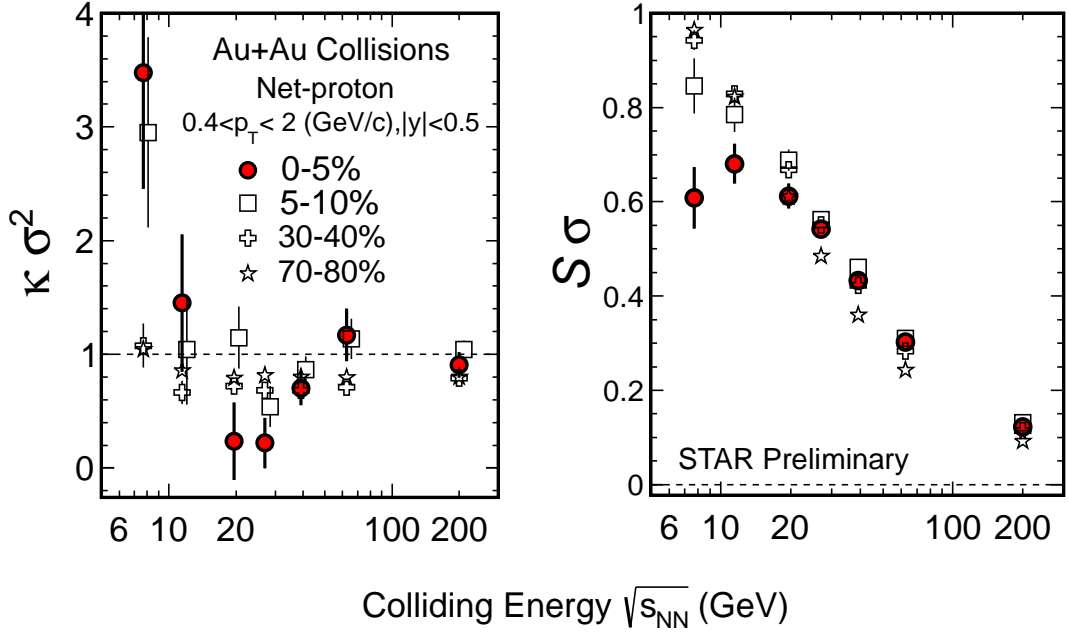}
%\vspace{-1.2cm}
\caption[]{(Color online) Energy dependence of efficiency corrected cumulant ratios $\kappa\sigma^{2}=C_{4}/C_{2}$ and $S\sigma=C_{3}/C_{2}$ of net-proton distributions in Au+Au collisions at different centralities ($0\sim5\%,5\sim10\%,30\sim40\%, 70\sim80\%$). } \label{fig:Energy_KV_SD}
%\end{center}
\end{figure}

In Fig. \ref{fig:Energy_KV_SD}, we present the energy dependence of efficiency-corrected cumulant ratios $\kappa\sigma^{2}=C_{4}/C_{2}$ and $S\sigma=C_{3}/C_{2}$ of net-proton distributions in Au+Au collisions at different centralities ($0\sim5\%,5\sim10\%,30\sim40\%, 70\sim80\%$). For peripheral ($70\sim80\%$) and mid-central ($30\sim40\%$) collisions, the {\KV} values are close to unity and the {\SD} show strong monotonic increase when the energy
decreases. For $0\sim5\%$ most-central collisions, the values of {\KV} are close to unity at energies above 39 GeV, while below 39 GeV,  they start to deviate from unity and show significant deviation below unity around 19.6 and 27 GeV. 
Finally, they shows a strong increase and stay above unity at 7.7 GeV. The  {\SD} at $0\sim5\%$ centrality bin shows a large drop at 7.7 GeV. One may note that we only have statistical errors shown in the figure, which are still large due to limited statistics. The systematical errors, which are dominated by the efficiency correction and the particle identification, are being studied.

\begin{figure}[htbp]
\begin{center}
%\hspace{-0.4in}
%\vspace{-0.8cm}
\includegraphics[scale=0.65]{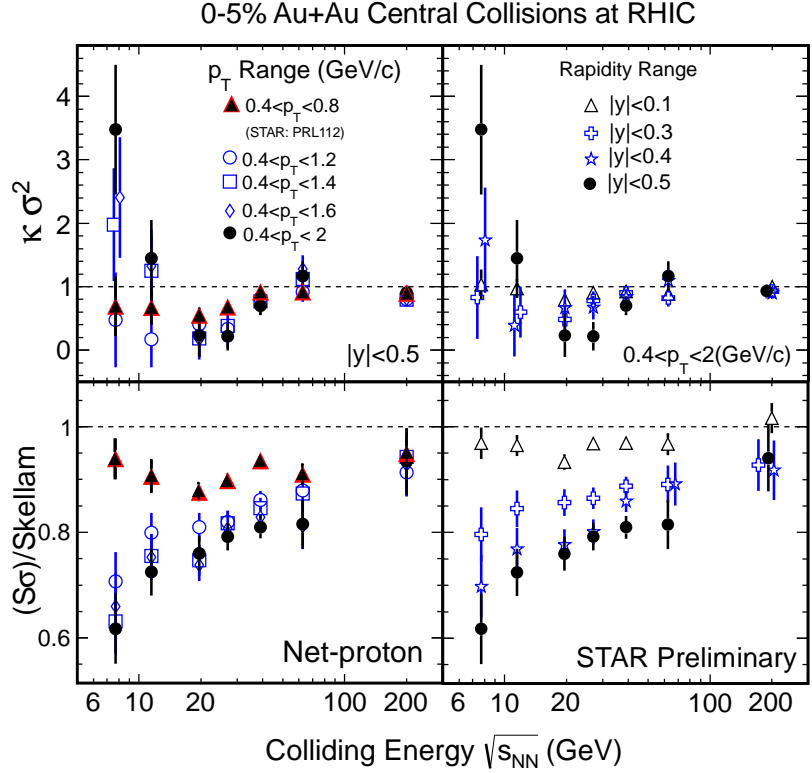}
%\vspace{-1.2cm}
\caption[]{(Color online) Energy dependence of efficiency corrected  $\kappa\sigma^{2}=C_{4}/C_{2}$ and $S\sigma/Skellam$ of net-proton distributions for different $p_{T}$ and rapidity ranges in $0\sim5\%$ most central Au+Au collisions. } \label{fig:PT_Y_Energy}
\end{center}
\end{figure}

Large acceptance is crucial for fluctuations of conserved quantities in heavy-ion collisions to probe the QCD phase transition and critical point. 
The signals for the phase transition and/or CP will be suppressed with small acceptance. In the Fig. \ref{fig:PT_Y_Energy}, we show the energy dependence of efficiency corrected  $\kappa\sigma^{2}=C_{4}/C_{2}$ and $S\sigma$/Skellam of net-proton distributions with various $p_{T}$ and rapidity range for $0\sim5\%$ most central Au+Au collisions. The Skellam baseline assumes the protons and anti-protons distribute as independent Poisson distributions. It is constructed from the efficiency-corrected mean values of the protons and anti-protons. It is expected to represent the thermal statistical fluctuations of the net-proton number~\cite{HRG_Karsch}. The {\KV} and {\SD}/Skellam are to be unity for Skellam baseline as well as in the Hadron Resonance Gas model.  In the two upper panels of Fig. \ref{fig:PT_Y_Energy}, when we gradually enlarge the $p_{T}$ or rapidity acceptance, the values of {\KV} show a small changes close to unity at energies above 39 GeV, while below 39 GeV, more pronounced structure is observed for a larger $p_{T}$ or rapidity acceptance.  In the two lower panels of Fig. \ref{fig:PT_Y_Energy}, when we enlarge the $p_{T}$ or rapidity acceptance, the {\SD}/Skellam shows strong suppression with respect to unity and monotonically decrease with energy. In contrast to {\KV}, the significantly increase above unity at 7.7 GeV is not observed in {\SD}/Skellam, but shows strong suppression below unity. The published results are shown as solid red triangles in the figure.

\begin{figure}[htbp]
\begin{center}
%\hspace{-0.4in}
%\vspace{-0.8cm}
\includegraphics[scale=0.35]{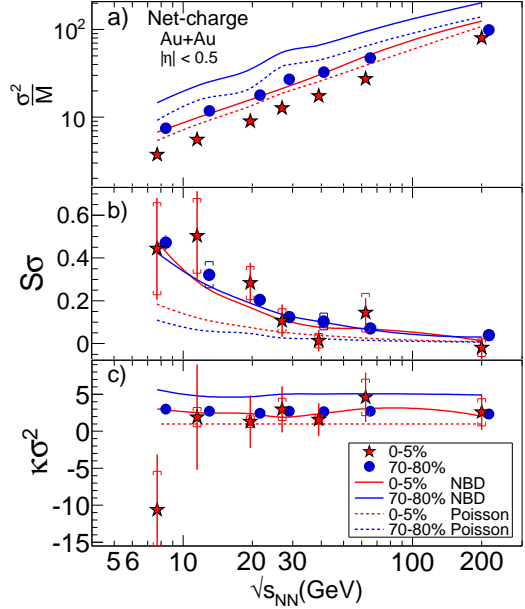}
%\vspace{-1.2cm}
\caption[]{(Color online) Energy dependence of moments of net-charge distributions for Au+Au collisions at RHIC BES energies.
The statistical and systematical error are shown in bars and brackets, respectively. } \label{fig:netcharge}
\end{center}
\end{figure}

The efficiency-corrected net-charge results are shown in Fig. \ref{fig:netcharge}. We did not observe non-monotonic behavior for {\SD} and {\KV} within current statistics for net-charge. The expectations from negative binomial distribution can better describe the data than the Poisson (Skellam) distribution. More statistics are needed for the net-charge analysis, especially at the lowest beam energies. The moments of net-charge
and net-proton distributions have been also used to extract the chemical freeze out parameters ($T$ and ${\mu_{B}}$) in heavy-ion collisions by comparing with the Lattice QCD or HRG model calculations~\cite{freezeout}.
We would like to discuss several observations: (1)  The STAR experiment has also carried out moment analysis for net-proton, net-charge, and net-kaon. Different measurements could be affected by kinematic cuts, resonance decays,  and other dynamical effects differently. In terms of searching for the QCD critical point,  careful studies are needed. (2) So far, the resonance decay effects are not eliminated in the experimental results of net-proton and net-charge moments. In principle, one can estimate the decay effects with theoretical calculations and/or models. Based on the Hadron Resonance Gas model calculation~\cite{HRG_baseline}, the decay effects for net-proton {\KV} is negligible  $\sim2\% $.  However, the model studies indicate the decay effects are large for net-charge moments~\cite{HRG_baseline}. (3) According to the analytical formula derived from the Delta theorem,  the statistical error of cumulants ($\Delta(C_n$)) are related to the width of the distribution as $\Delta(C_n)$ $\sim$ O($\sigma^{n}$)~\cite{Delta_theory,Unified_Errors}. Thus, for the same number of events, the wider the distribution is, the larger the statistical errors. This explains why the net-charge moments have larger statistical errors than those of net-proton, as the former has a much wider distribution. (4) Theoretical calculations show that the coupling strength of pions to critical fluctuations is smaller than of protons/anti-protons. Thus, one may not be able to see strong CP signals in the fluctuations of net-charges (dominated by pions)~\cite{ratioCumulant, privateCom}. 

In the year 2018, the second phase of the beam energy scan program (BES-II) at RHIC will commence. During BES-II, we plan to fine tune the beam energies below 20 GeV and  accumulate more statistics as the luminosity can be increased about $3\sim10$ times via stochastic electron cooling. Several sub-detector upgrades for the STAR experiment are ongoing and will be ready when BES-II starts. Two of the upgrades will benefit the moment analysis at STAR.  One is the inner TPC (iTPC) upgrade, which will improve the tracking efficiency and enlarge the TPC acceptance and the other is installation of a new detector, the so-called Event Plane Detector (EPD), which is a forward detector used to determine the event plane for Au+Au collisions. It can provide centrality determination with particles far away from the collision region.

\section{Summary}
We have reviewed the analysis techniques used in the moment analysis and presented the energy dependence of moments of net-proton and net-charge distributions in Au+Au collisions measured by the STAR experiment in the first phase of the RHIC BES program. The upper transverse momentum ($p_{T}$) limit for proton and anti-proton is extended up to 2 GeV/c with the time of flight (ToF) detector for particle identification. To obtain efficiency-corrected moments, a unified method for implementing phase space efficiency correction and calculating statistical errors is applied in the net-proton moment analysis. Generally, the $p_{T}$ and rapidity acceptance dependence study for the moments of the net-proton distributions shows that the larger the acceptance is, the larger deviations from unity. The most pronounced structure is observed in the energy dependence of {\KV} of net-proton distributions from the $0\sim5\%$ most central collisions within $0.4<p_{T}<2$ GeV/c and at mid-rapidity $|y|<0.5$. At energies above 39 GeV,  the values of {\KV} are close to unity and for energies below 39 GeV, it shows significant deviation below unity around 19.6 and 27 GeV, then large increase above unity is observed at 7.7 GeV. Finally, careful model and/or theoretical studies are needed to understand the effects of non-critical contributions associated with the heavy-ion collision dynamics.

\section*{Acknowledgement}
The work was supported in part by the MoST of China 973-Project No. 2015CB856901, NSFC under grant No. 11205067, 11221504 and 11228513.

\bibliography{CPOD2014_proceedings_XFLUO}
\bibliographystyle{unsrt}

\end{document}